\documentclass[12pt]{article}

\usepackage{epsfig}
\usepackage{graphicx}
\usepackage{amssymb,amsmath}
\usepackage{hyperref}
\hypersetup{
    colorlinks=true,
    linkcolor=blue,
    citecolor=red,
}

\begin{document}

\begin{center}
\Large{\bf Generalized Uncertainty Principle effects in the
Ho\v{r}ava-Lifshitz quantum theory of gravity} \vspace{0.5cm}

\large  H. Garc\'{\i}a-Compe\'an\footnote{e-mail address: {\tt
compean@fis.cinvestav.mx}}, D. Mata-Pacheco\footnote{e-mail
address: {\tt dmata@fis.cinvestav.mx}}

\vspace{0.3cm}

{\small \em Departamento de F\'{\i}sica, Centro de
Investigaci\'on y de Estudios Avanzados del IPN}\\
{\small\em P.O. Box 14-740, CP. 07000, Ciudad de M\'exico, M\'exico}\\

\vspace*{1.5cm}
\end{center}

\begin{abstract}
The Wheeler-DeWitt equation for a Kantowski-Sachs metric in
Ho\v{r}ava-Lif\-shitz gravity with a set of coordinates in
minisuperspace that obey a generalized uncertainty principle is
studied. We first study the equation coming from a set of
coordinates that obey the usual uncertainty principle and find
analytic solutions in the infrared as well as a particular
ultraviolet limit that allows us to find the solution found in
Ho\v{r}ava-Lifshitz gravity with projectability and with detailed
balance but now as an approximation of the theory without detailed
balance. We then consider the coordinates that obey the generalized
uncertainty principle by modifying the previous equation using the
relations between both sets of coordinates. We describe two possible
ways of obtaining the Wheeler-DeWitt equation. One of them is useful to
present the general equation but it is found to be very difficult to
solve. Then we use the other proposal to study the limiting cases
considered before, that is, the infrared limit that can be compared
to the equation obtained by using general relativity and the
particular ultraviolet limit. For the second limit we use a
ultraviolet approximation and then solve analytically the resulting
equation. We find that and oscillatory behaviour is possible in this context
but it is not a general feature for any values of the parameters
involved. \vskip 1truecm

\end{abstract}

\bigskip

\newpage

\section{Introduction}
\label{S-Intro}

Although we do not yet have a complete description of a theory of
quantum gravity, there have been enormous efforts from very
different points of view, that has brought us closer to an
understanding of the gravitational phenomena in a quantum regime.
For example, there are now some features that are expected to be
found in such a theory. One of the main features that we expect from
a quantum theory of gravity is the existence of a minimal measurable
length. This idea has arisen in a number of approaches to
quantum gravity ranging from string theory \cite{STLength,STLength2}
to general thought experiments of black holes physics
\cite{Maggiore:1993rv,Scardigli:1999jh} to name a few (see for
example \cite{Garay:1994en} for a more thorough discussion). One of
the features of the hypothesis of the existence of a minimal length
is a modification of the Heisenberg uncertainty principle and thus a generalization of the Heisenberg algebra, known as the
Generalized Uncertainty Principle (GUP). This proposal has been
extensively studied in the literature from the purely theoretical
point of view such as in Refs. \cite{Kempf:1994su,Anacleto:2015mma,Anacleto:2015rlz}, as well as from
attempts to constraint the parameters involved using well measured
quantities from both the classical as well as the quantum point of
view, for example see Refs.
\cite{Scardigli:2016ubl,Scardigli:2016pjs,Lambiase:2017adh,Vagenas:2017fwa,Bosso:2017hoq,Demir:2018akw,Bosso:2018ckz,Fu:2021zrd}, and even on experimental grounds such as Ref. \cite{Bushev:2019zvw}.
These effects are expected to be relevant on systems at very high
energies (or short distances) and therefore they are expected to be of first importance
in the early universe and quantum effects of black holes. One of the main tools to our disposal to
approach the quantum behaviour of gravity is the well known tools of
canonical quantization that leads to the Wheeler-DeWitt (WDW)
equation \cite{Arnowitt:1962hi,Wheeler,DeWitt} and therefore it is
natural to study the effects of a GUP in this framework. In fact,
recently the effects of considering a GUP to the coordinates in the
minisuperspace for a Kantowsi-Sachs metric in the context of the WDW
equation was introduced in \cite{GUP}. For a short overview on this topic the reader can consult Ref. \cite{Garcia-Compean:2021wgv}. It has also been studied recently a comparison between considering a GUP and Polymer Quantum Mechanics in a semiclassical as well as quantum approaches in \cite{Barca:2021epy}.

On the other hand, it is well known that General Relativity (GR) is
a non-renorma\-liza\-ble theory and thus looking for quantum aspects
of gravity using this theory has a limiting nature. Apart from the
search of a full theory of quantum gravity, there has been very
interesting approaches that seek to generalize GR by modifying it so
it has a better behaviour in the ultraviolet. One of the most
important proposals on this regard is the Ho\v{r}ava-Lifshitz (HL)
theory of gravity \cite{Horava:2009uw} (for some recent reviews, see
\cite{Weinfurtner:2010hz,Sotiriou:2010wn,Wang:2017brl,Mukohyama:2010xz}
and references therein). This theory employs an anisotropic scaling
of space and time variables, inspired by  the Lifshitz scaling of the
condensed matter physics, which breaks the Lorentz symmetry of GR and
introduces spatial  higher-derivative terms to the action of gravity
in order to achieve a power counting renormalizable theory. Since
this theory represents an improvement in the ultraviolet behaviour of
GR, the theory has been studied extensively in the context of
canonical quantization, that is regarding the WDW equation, for
example see Refs.
\cite{Bertolami:2011ka,Christodoulakis:2011np,Pitelli:2012sj,Vakili:2013wc,Obregon:2012bt,Benedetti:2014dra,Cordero:2017egl}.
Although the non-projectable version of the theory has advantages
over the projectable version in particular regarding the infrared
limit of the theory, it is also well known that when working with
the WDW equation both approaches are completely equivalent and the
infrared instability is not detected at this level and therefore in this work we will use the projectable
theory. However in the
most general case, this perturbative instability leads to the
existence of an additional degree of freedom, which does not allow
to recover GR (see for instance, \cite{Cordero:2017egl} for an
explanation).

Since HL theory is expected to produce a better quantum behaviour
than GR and the GUP is an expected feature of quantum gravity, it is
natural to consider both of them in the same approach. A possible
connection between these two proposals has been previously studied
in Refs. \cite{Myung:2009ur,Myung:2009gv}. However, in the present
article we are going to consider both proposals as independent. We
are going to study the effects of considering a GUP in the WDW
equation obtained in the context of HL gravity for a Kantowski-Sachs
(KS) metric, considering therefore two contributions in the
ultraviolet quantum gravity behaviour in the WDW equation. Thus we
will generalize the proposal used for GR in \cite{GUP}.

The present work is organized as follows. In Section
\ref{S-KSNormal} we are going to study the standard WDW equation
(that is with coordinates that obey the usual uncertainty principle)
for a Kantowski-Sachs metric in HL gravity. We will particularly
focus on two regimes, the infrared (IR) limit and a special case in the
ultraviolet (UV) limit that leads to an analytic solution. In
Section \ref{S-KSGUP} we will consider a new set of variables that
obeys a GUP and we will show how the WDW equation is obtained in
this context, we will present two equivalent ways to deduce such
equation. Since the resulting equation is far quite difficult to
solve, in Section \ref{S-KSIRUV} we will consider the two simple
cases considered previously, that is, the IR limit and the
particular UV limit. The IR limit can be compared to the case
considered in \cite{GUP} using GR. In the UV limit we look for
analytic solutions and we present some figures where we plot 
the resulting behaviour. Finally, in Section \ref{S-FinalR} we present
our conclusions and final remarks.

\section{Wheeler-DeWitt equation in Ho\v{r}ava-Lifshitz gravity}
\label{S-KSNormal} In this section we are going to study the WDW
equation coming from a Kantowski-Sachs metric in HL gravity with the
usual uncertainty relation. We will study its UV and IR limits,
taking special interest in cases in which analytic solutions can be
obtained.

We begin by considering the gravitational action in projectable HL
gravity without detailed balance. This action can be written as
\cite{Bertolami:2011ka,Sotiriou:2009gy,Sotiriou:2009bx}
    \begin{multline}\label{ActionHL}
        S=\frac{M^2_{pl}}{2}\int_{M}d^3xdtN\sqrt{h}\left[K_{ij}K^{ij}-\lambda K^2-\Lambda M^2_{pl}+R-\frac{1}{M^2_{pl}}\left(g_{2}R^2+g_{3}R_{ij}R^{ij}\right)\right. \\ \left. -\frac{1}{M^4_{pl}}\left(g_{4}R^{3}+g_{5}RR^{i}_{j}R^{j}_{i}+g_{6}R^{i}_{j}R^{j}_{k}R^{k}_{i}+g_{7}R\nabla^{2}R+g_{8}\nabla_{i}R_{jk}\nabla^{i}R^{jk}\right)\right]\\ + M^2_{pl}\int_{\partial M}d^3x\sqrt{h}K,
    \end{multline}
where $N$ is the lapse function, $\lambda$ is a running parameter proper of the Ho\v{r}ava-Lifshitz theory, which is associated to the fact that the introduction of an anisotropic scaling $z$ leads to have a reduced symmetry which is the diffeomorphisms that preserve a preferred foliation ${\cal F}$Diff, $g_{n}$ are dimensionless couplings, $M_{pl}$ is the Planck
mass, $K_{ij}$ is the extrinsic curvature and $h_{ij}$ is the
induced 3-metric on the three-dimensional leave of the foliation.

We will consider the Kantowski-Sachs metric, which describes an
homogeneous but anisotropic metric that can be used to describe an
anisotropic cosmological model or the interior of a Schwarzschild
black hole. This metric can be written in the parametrization form
proposed by Misner \cite{Misner} as
\begin{equation}\label{KSMetric}
ds^2=-N^2(t)dt^2+e^{2\sqrt{3}\beta}dr^2+e^{-2\sqrt{3}(\beta+\Omega)}\left(d\theta^2+\sin^2\theta
d\phi^2\right),
\end{equation}
where $\beta$ and $\Omega$ are real functions of the cosmological time parameter. These are the coordinates of the minisuperspace and they are associated to the anisotropic directions in the cosmological metric.  For this metric, the above action (\ref{ActionHL}) reads
    \begin{multline}\label{ActionKSHL}
        S=V_{0}\int dt N\left\{\left[\frac{3}{N^2}(3-\lambda)\dot{\beta}^2+\frac{12}{N^2}(1-\lambda)\dot{\beta}\dot{\Omega}+\frac{6}{N^2}(1-2\lambda)\dot{\Omega}^2\right]e^{-\sqrt{3}(\beta+2\Omega)}\right. \\ \left. +2e^{\sqrt{3}\beta}\left(1-\Lambda
        e^{-2\sqrt{3}(\beta+\Omega)}\right)-g_{r}e^{\sqrt{3}(3\beta+2\Omega)}-g_{s}e^{\sqrt{3}(5\beta+4\Omega)}\right\},
    \end{multline}
where we have defined
    \begin{equation}
        g_{r}=  2g_{2}+g_{3} , \hspace{1cm} g_{s}=
        2g_{4}+g_{5}+\frac{g_{6}}{2}.
    \end{equation}
The volume of the spatial slice is in this case given by
    \begin{equation}
        V_{0}=\int\sin\theta d^3x,
    \end{equation}
and we have chosen units such that $\frac{M^2_{pl}}{2}=1$ with
$\hbar=c =1$. We note that if we consider a properly compactified
spatial slice then $V_{0}$ would be a finite factor. However, since
it only plays the role of a global multiplicative factor in the
action we will ignore it hereafter. Following the standard
procedure, we can compute the hamiltonian defined by
    \begin{equation}\label{DefHam}
        H=P_{\beta}\dot{\beta}+P_{\Omega}\dot{\Omega}+P_{N}\dot{N}-\mathcal{L},
    \end{equation}
where $\mathcal{L}$ is the lagrangian appearing in the action and
$P_{\beta},P_{\Omega}, P_{N}$ are canonical momenta of the
corresponding variables. Since the system is reparametrization invariant, the hamiltonian vanishes and therefore we are led to a hamiltonian constraint in the form
    \begin{multline}\label{HamiltonianConstraintO}
        H=\frac{N}{12}\frac{e^{\sqrt{3}(\beta+2\Omega)}}{3\lambda-1}\left[(2\lambda-1)P^2_{\beta}-2(\lambda-1)P_{\beta}P_{\Omega}+\frac{1}{2}(\lambda-3)P^2_{\Omega}\right. \\ \left. -24(3\lambda-1)e^{-2\sqrt{3}\Omega}\left(1-\Lambda e^{-2\sqrt{3}(\beta+\Omega)}\right)+12g_{r}(3\lambda-1)e^{2\sqrt{3}\beta}\right. \\ \left.
        +12g_{s}(3\lambda-1)e^{2\sqrt{3}(2\beta+\Omega)}\right]\simeq0 ,
    \end{multline}
which holds globally since in the projectable version of the theory the lapse function is taken to be only a function of time and the dependence on the spatial variables has been integrated leading to a global multiplicative factor. If we would have considered the non-projectable version of the theory, we would have obtained the same constraint but that would be valid only locally. However, since all the fields depend only on the time variable, they are both equivalent always. The first five terms of the hamiltonian corresponds basically to the
model of KS in the context of the Ho\v{r}ava-Lifshitz gravity with
detailed balance conditions studied in \cite{Obregon:2012bt}. The
additional term is a higher-correction in the radius of the KS model
of the form $e^{4\sqrt{3}\beta +2 \sqrt{3}\Omega}$ which is a
manifestation of the cubic terms in $R$. Just as it was described in
Refs.
\cite{Bertolami:2011ka,Pitelli:2012sj,Sotiriou:2009gy,Sotiriou:2009bx}
the projectable model without detailed balanced discussed in the
present section reduces consistently to the one with detailed
balance. The standard WDW equation for this metric is obtained after
canonical quantization of the hamiltonian constraint, that is after
considering
    \begin{equation}\label{MomentaOrig}
    P_{\beta}=-i\frac{\partial}{\partial\beta} , \hspace{1cm} P_{\Omega}=-i\frac{\partial}{\partial\Omega},
    \end{equation}
and the standard commutation relations
    \begin{equation}\label{UPO}
        \left[\Omega,P_{\Omega}\right]=i, \hspace{0.5cm} \left[\beta,P_{\Omega}\right]=0 , \hspace{0.5cm} \left[\Omega,P_{\beta}\right]=0 , \hspace{0.5cm} \left[\beta,P_{\beta}\right]=i.
    \end{equation}
The general WDW equation obtained in this way is given by
    \begin{multline}\label{WDWO}
        \left[\frac{1}{2}(\lambda-3)\frac{\partial^2}{\partial\Omega^2}-2(\lambda-1)\frac{\partial}{\partial\Omega}\frac{\partial}{\partial\beta}+(2\lambda-1)\frac{\partial^2}{\partial\beta^2}\right. \\ \left. +24(3\lambda-1)e^{-2\sqrt{3}\Omega}\left(1-\Lambda e^{-2\sqrt{3}(\beta+\Omega)}-\frac{g_{r}}{2}e^{2\sqrt{3}(\beta+\Omega)}-\frac{g_{s}}{2}e^{4\sqrt{3}(\beta+\Omega)}\right)\right]\Psi(\beta,\Omega)=0.
    \end{multline}
Since this equation is very difficult to solve in general, two
particular cases can be considered in order to give exact solutions.

Let us remember that in the Misner parametrization we have
$t=e^{-\sqrt{3}(\beta+\Omega)}$. Therefore, the IR limit is obtained
when $t>>1$ and $\lambda\to 1$. In this case we have very small
curvature with a corresponding very large curvature radius with
respect to the Planck length $R_{c}>>L_{pl}$. In this case the
biggest contribution comes from the exponentials with negative
signs. Taking a vanishing cosmological constant we obtain in this
limit
    \begin{equation}\label{WDWOGR}
        \left[-\frac{\partial^2}{\partial\Omega^2}+\frac{\partial^2}{\partial\beta^2}+48e^{-2\sqrt{3}\Omega}\right]\Psi(\Omega,\beta)=0,
    \end{equation}
which corresponds to the well known model of GR without a
cosmological constant. The solutions are of the form
    \begin{equation}\label{SolWDWOGR}
        \Psi^{\pm}_{\nu}(\Omega,\beta)=e^{\pm i\nu\sqrt{3}\beta}K_{i\nu}\left(4e^{-\sqrt{3}\Omega}\right),
    \end{equation}
where $K_{i\nu}$ are the modified Bessel functions of the second kind.

\begin{figure}[h!]
	\centering
	\includegraphics[width=0.50\textwidth]{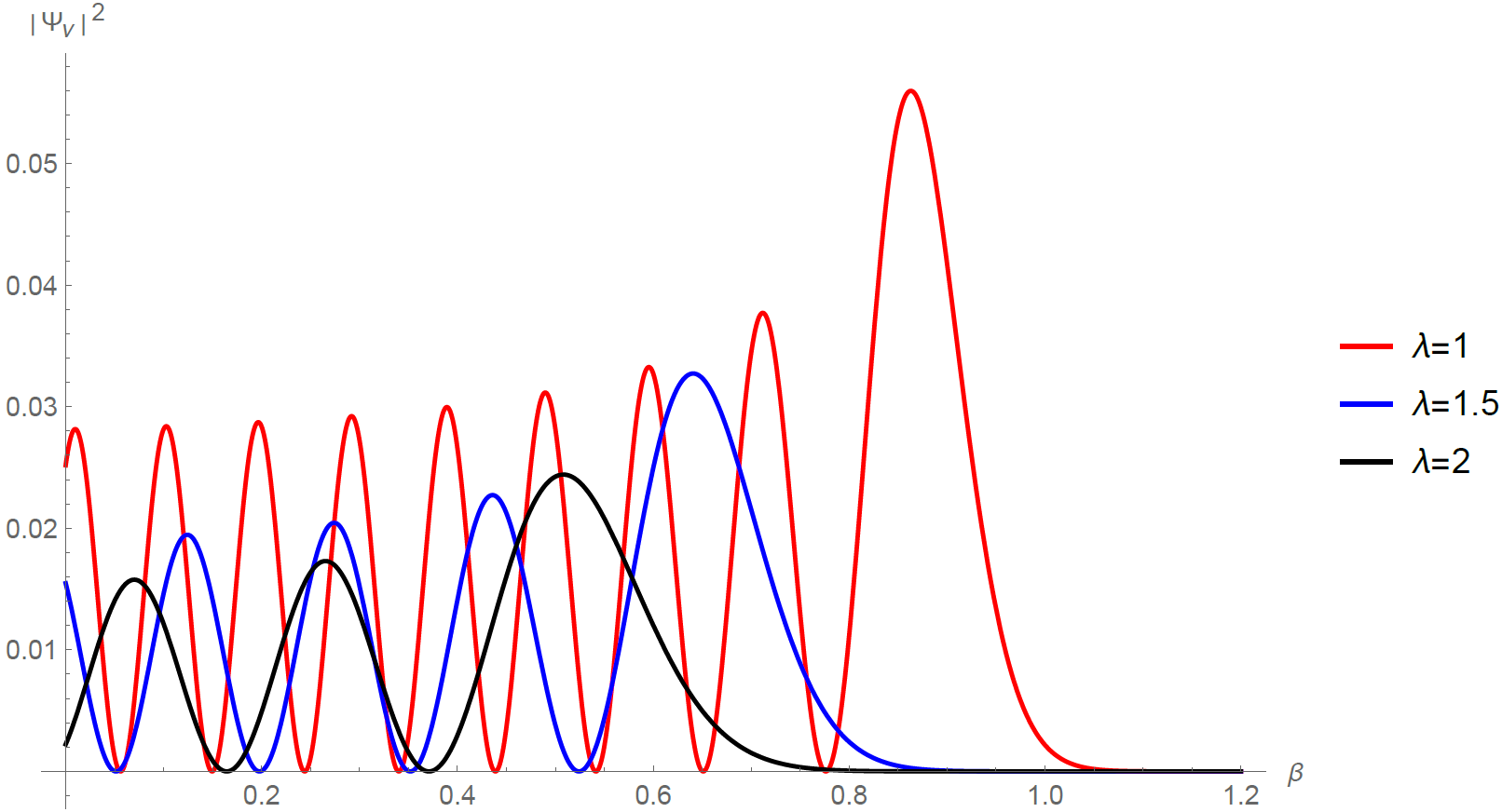}
	\caption{Absolute value squared of the wave functional $|\Psi_{\nu}|^2$ choosing $\nu=20i$ for $\lambda=1$ (red curve), $\lambda=1.5$ (blue curve) and $\lambda=2$ (black curve). }
	\label{WGUP}
\end{figure}

We can also consider the opposite case, that is the UV limit. This
limit is achieved when $t<<1$. In this case we have a huge curvature
with a corresponding very small curvature radius which may be of the
magnitude order of the Planck length. Then the exponentials with
positive signs in (\ref{HamiltonianConstraintO}) are the main
contribution to the equation. In order to simplify the WDW equation
we will consider however a particular case in which $g_{r}=2$ and
$g_{s}=0$, in this case we obtain
    \begin{equation}\label{WDWOUV}
    \left[\frac{\lambda-3}{2}\frac{\partial^2}{\partial\Omega^2}-2(\lambda-1)\frac{\partial}{\partial\Omega}\frac{\partial}{\partial\beta}+(2\lambda-1)\frac{\partial^2}{\partial\beta^2}-3\mu^2(2\lambda-1)e^{2\sqrt{3}\beta}\right]\Psi(\Omega,\beta)=0,
    \end{equation}
with $\mu^2=8\frac{3\lambda-1}{2\lambda-1}$.  This last WDW equation
(\ref{WDWOUV}) was obtained in Ref. \cite{Obregon:2012bt} in the
context of the Ho\v{r}ava-Lifshitz gravity with projectability and
with detailed balance. However we have shown that Eq. (\ref{WDWOUV})
can also be obtained as an approximation of the theory without
detailed balance. This equation has analytical solutions which can
be expressed in a normalized form as
    \begin{multline}\label{SolWDWUV}
        \Psi_{\nu}(\beta,\Omega)=\frac{3^{1/4}}{\pi}\frac{\sqrt{3\lambda-1}}{\sqrt{2}(2\lambda-1)}\sqrt{\sinh\left(\frac{\pi\sqrt{3\lambda-1}|\nu|}{\sqrt{2}(2\lambda-1)}\right)}e^{\pm \sqrt{3}\nu\Omega}\left[\mu e^{\sqrt{3}\beta}\right]^{\frac{\lambda-1}{2\lambda-1}\nu}\\ \times K_{\frac{\sqrt{3\lambda-1}}{\sqrt{2}(2\lambda-1)}\nu}\left(\mu e^{\sqrt{3}\beta}\right),
    \end{multline}
where $\nu$ is restricted to be a purely imaginary number. In order to visualize the behaviour of this solution we plot the squared absolute value of the wave functional for different values of $\lambda$ and show the results in Figure \ref{WGUP}. We note that since in this case $\nu$ is a purely imaginary number, the dependence on $\Omega$ to the absolute value squared of the wave functional vanishes and therefore we only plot against the variable $\beta$. The behaviour obtained corresponds to a region of oscillations with increasing height that reaches a maximum peak and then it stops oscillating and decreases. This is the behaviour obtained in general for any value of $\lambda$ in the region of interest. We also note that as $\lambda$ increases, the height of the oscillations in general decreases and the region of oscillatory behaviour is also reduced.

\section{Ho\v{r}ava-Lifshitz  gravity with GUP}
\label{S-KSGUP}

Now that we have considered the WDW equation coming from the
Kantowski-Sachs metric in HL gravity with the standard commutation
relations, we now move on to study the changes made by considering
the introduction of variables that obey a GUP. In particular we are
going to use the relations  inspired by \cite{Kempf:1994su}, where a
commutation relation that leads to a nonzero minimum uncertainty in
the position variable is considered and that was used in \cite{GUP}
in the context of the WDW equation in GR. To be more precise we
consider that the variables in minisuperspace obey the commutation
relation
    \begin{equation}\label{RelationGUP}
        [q_{i},p_{j}]=i  \delta_{ij}\left(1+\gamma^2p^2\right),
    \end{equation}
where $q_{1}=\Omega$, $q_{2}=\beta$, $p_{1}=P_{\Omega}$,
$p_{\beta}=P_{\beta}$, $\gamma$ is a small parameter with units of
inverse momentum and $p^2=p^{k}p_{k}$ is computed using the metric
in superspace. We also introduce coordinates $q'_{i}$ that obey the
usual commutation relations, that is $[q'_{i},p_{j}]=i\delta_{ij}$.
In position space both sets of coordinates can be related as
    \begin{equation}\label{RelationsCoordinates}
        q_{i}=q'_{i}\left(1+\gamma^2p^2\right).
    \end{equation}
We can also choose the following representation for the momentum operators
    \begin{equation}\label{MomOperatorsGUP}
        P_{\Omega}=-i \frac{\partial}{\partial\Omega'} , \hspace{1cm} P_{\beta}=-i \frac{\partial}{\partial\beta'}.
    \end{equation}
We also note from (\ref{WDWO}) that in this case
    \begin{equation}\label{PS}
        p^2=(2\lambda-1)P^2_{\beta}-2(\lambda-1)P_{\beta}P_{\Omega}+\frac{1}{2}(\lambda-3)P^2_{\Omega}.
    \end{equation}

In order to obtain a WDW equation with this setup we have to rewrite
each term in equation (\ref{HamiltonianConstraintO}) in terms of the
prime coordinates. To be able to rewrite all the exponential terms
we are going to use repeatedly the Zassenhaus formula which states
the following
    \begin{equation}\label{ZSFormula}
        e^{A+B}=e^{A}e^{B}e^{-\frac{1}{2}[A,B]}e^{\frac{1}{6}\left([A,[A,B]]+2[B,[A,B]]\right)}
        \cdots
    \end{equation}
where $\cdots$ denotes terms with commutators involving more than 3
operators.

To begin with we will consider the first term contributing to the
potential in (\ref{WDWO}), using (\ref{RelationsCoordinates}) we can
write it as
    \begin{equation}\label{ExpansionOmegaM1}
        \exp[-2\sqrt{3}\Omega]=\exp\left[-2\sqrt{3}\Omega'-2\sqrt{3}\gamma^2\Omega'p^2
        \right].
    \end{equation}
Therefore, using the Zassenhaus formula (\ref{ZSFormula}) with
$A=-2\sqrt{3}\Omega'$ and $B=-2\sqrt{3}\gamma^2\Omega'p^2$ and
considering only up to second order terms in $\gamma$ we obtain
    \begin{multline}\label{ExpansionOmegaM}
        \exp[-2\sqrt{3}\Omega]\simeq \exp\left[-2\sqrt{3}\left\{1-2\gamma^2(\lambda-3)\right\}\Omega'\right]\exp\left[-2\sqrt{3}\gamma^2\Omega'p^2\right]
        \\ \times \exp\left[-6i\gamma^2\Omega'\left\{(\lambda-3)P_{\Omega}-2(\lambda-1)P_{\beta}\right\}\right].
    \end{multline}
Carrying out the same procedure we obtain for the remaining terms
    \begin{multline}\label{ExpansionBetaP}
    \exp\left[2\sqrt{3}\beta\right]\simeq\exp\left[2\sqrt{3}\left\{1-4\gamma^2(2\lambda-1)\right\}\beta'\right]\exp\left[2\sqrt{3}\gamma^2\beta'p^{2}\right]\\ \times \exp\left[-12i\gamma^2\beta'\left\{(2\lambda-1)P_{\beta}-(\lambda-1)P_{\Omega}\right\}\right],
    \end{multline}
    \begin{multline}\label{ExpansionBOM}
    \exp\left[-2\sqrt{3}(\beta+2\Omega)\right]\simeq \exp\left[-2\sqrt{3}(1+12\gamma^2)(\beta'+2\Omega')\right]\exp\left[-2\sqrt{3}\gamma^2(\beta'+2\Omega')p^2\right] \\  \times \exp\left[-12i\gamma^2(\beta'+2\Omega')(P_{\beta}-2P_{\Omega})\right],
    \end{multline}
    \begin{multline}\label{ExpansionBOP}
    \exp\left[2\sqrt{3}(2\beta+\Omega)\right] \simeq \exp\left[2\sqrt{3}\left\{1-6\gamma^2(3\lambda-1)\right\}(2\beta'+\Omega')\right]\exp\left[2\sqrt{3}\gamma^2(2\beta'+\Omega')p^2\right] \\  \times \exp\left[-6i\gamma^2(3\lambda-1)(2\beta'+\Omega')(2P_{\beta}-P_{\Omega})\right].
    \end{multline}

Now that we have expressed all the factors in terms of prime
coordinates we proceed to study its behaviour when they are applied
to the wave functional. Let us study first the term in
(\ref{ExpansionOmegaM}). First of all up to second order in $\gamma$
and momentum we obtain
        \begin{multline}\label{AppWF11}
        e^{-2\sqrt{3}\Omega}\Psi(\Omega,\beta)\simeq\exp\left[-2\sqrt{3}\left\{1-2\gamma^2(\lambda-3)\right\}\Omega'\right]\exp\left[-2\sqrt{3}\gamma^2\Omega'p^2\right]\\ \times \exp\left[-6i\gamma^2\Omega'\left\{(\lambda-3)P_{\Omega}-2(\lambda-1)P_{\beta}\right\}\right]\Psi(\Omega',\beta'),
        \end{multline}
Then, we have two options to study the result of applying the
exponential terms with linear momenta to the wave functional. The
first one is to note that up to second order in $\gamma$ we can
write
    \begin{multline}\label{Expansion}
    \exp\left[-6i\gamma^2\Omega'\left\{(\lambda-3)P_{\Omega}-2(\lambda-1)P_{\beta}\right\}\right]\\=\exp\left[12i\gamma^2(\lambda-1)\Omega'P_{\beta}\right]\exp\left[-6i\gamma^2(\lambda-3)\Omega'P_{\Omega}\right],
    \end{multline}
and we note that by defining $\Omega'=e^{y}$ and using
(\ref{MomOperatorsGUP}) we have
    \begin{equation}
        \Omega'P_{\Omega}=-i \frac{\partial}{\partial y}.
    \end{equation}
Consequently the last term in (\ref{Expansion}) acts as a
translation operator for the variable $y$, that corresponds to a
scaling of the variable $\Omega'$. Thus we obtain
    \begin{equation}\label{ScalingDef}
        \exp\left[-6i\gamma^2(\lambda-3)\Omega'P_{\Omega}\right]\Psi(\Omega',\beta')=\Psi(e^{-6\gamma^2(\lambda-3)}\Omega',\beta').
    \end{equation}
Moreover, if we expand the exponential in the first term of
(\ref{Expansion}) as a power series and keep it only up to second
order in $\gamma$ we finally obtain for the first option
    \begin{multline}\label{FirstOptionExp}
        \exp\bigg\{-6i\gamma^2\Omega'\left((\lambda-3)P_{\Omega}-2(\lambda-1)P_{\beta}\right)\bigg\}\Psi(\Omega',\beta') \\ \simeq \left[1+12i\gamma^2\Omega'(\lambda-1)P_{\beta}\right]\Psi(e^{-6\gamma^2(\lambda-3)}\Omega',\beta').
    \end{multline}
Therefore, substituting it back into (\ref{AppWF11}) and expanding
as well the second exponential in the same way we obtain
\begin{multline} \label{AppWF1}
e^{-2\sqrt{3}\Omega}\Psi(\Omega,\beta)\simeq
e^{-2\sqrt{3}[1-2\gamma^2(\lambda-3)]\Omega'}
\bigg\{1+12i\gamma^2\Omega'(\lambda-1)P_{\beta} \\
 -2\sqrt{3}\gamma^2\Omega'p^2 \bigg\}
\Psi(e^{-6\gamma^2(\lambda-3)}\Omega',\beta').
\end{multline}

On the other hand, instead of interpreting the second term in
(\ref{Expansion}) as an scaling on $\Omega'$, as in
(\ref{ScalingDef}), we could also expand the exponential and keep
only up to second order in $\gamma$, therefore we have for the
second option
\begin{multline}\label{AppWF2}
e^{-2\sqrt{3}\Omega}\Psi(\Omega,\beta)\simeq\exp^{-2\sqrt{3}\big[1-2\gamma^2(\lambda-3)\big]\Omega'}\bigg\{1-6i\gamma^2\Omega'\big[(\lambda-3)P_{\Omega}-2(\lambda-1)P_{\beta}\big]
\\ -2\sqrt{3}\gamma^2\Omega'p^2\bigg\}\Psi(\Omega',\beta').
\end{multline}
We note that both options are equivalent at the level of
approximation used, that is, up to second order in $\gamma$ and
momenta. Therefore we can use each of the options when we find it
more useful.

We also note that the arguments used to get (\ref{ScalingDef}) are
applicable to any term that contains a factor of the form a product
of a coordinate times its momentum within an exponential. Therefore
any of the terms (\ref{ExpansionBetaP})-(\ref{ExpansionBOP}) have
also two possible expansions similarly to (\ref{AppWF1}) and
(\ref{AppWF2}), and we will use any of them when we find it more
useful.

Now that we have explained the general way in which any of the terms
contributing to the WDW equation (\ref{WDWO}) are expressed in terms
of prime coordinates and how they act on the corresponding wave
functional. Thus we obtain the general WDW equation that takes into
account the new commutation relations (\ref{RelationGUP}). Taking in
all cases the second option of the form (\ref{AppWF2}) for
simplicity we obtain the equation
    \begin{multline}\label{WDWGUPGeneral}
    \bigg\{\left[1+\gamma^2F(\Omega',\beta')\right]\left[(2\lambda-1)P^2_{\beta}-2(\lambda-1)P_{\beta}P_{\Omega}+\frac{1}{2}(\lambda-3)P^2_{\Omega}\right] \\  +i\gamma^2G(\Omega',\beta',P_{\Omega},P_{\beta}) +\widetilde{V}(\Omega',\beta')\bigg\}\Psi(\Omega',\beta')=0,
    \end{multline}
where we have defined
    \begin{multline}
    F(\Omega',\beta')=48\sqrt{3}(3\lambda-1)\Omega'e^{-2\sqrt{3}\left[1-2\gamma^2(\lambda-3)\right]\Omega'}\\-48\sqrt{3}(3\lambda-1)\Lambda(\beta'+2\Omega')e^{-2\sqrt{3}(1+12\gamma^2)(\beta'+2\Omega')} \\ +24\sqrt{3}g_{r}(3\lambda-1)\beta'e^{2\sqrt{3}\left[1-4\gamma^2(2\lambda-1)\right]\beta'}\\+24\sqrt{3}g_{s}(3\lambda-1)(2\beta'+\Omega')e^{2\sqrt{3}\left[1-6\gamma^2(3\lambda-1)\right](2\beta'+\Omega')},
    \end{multline}
    \begin{multline}
    G(\Omega',\beta',P_{\Omega},P_{\beta})=144(3\lambda-1)\Omega'\left[(\lambda-3)P_{\Omega}-2(\lambda-1)P_{\beta}\right]e^{-2\sqrt{3}\left[1-2\gamma^2(\lambda-3)\right]\Omega'}\\-288(3\lambda-1)\Lambda(\beta'+2\Omega')(P_{\beta}-2P_{\Omega})e^{-2\sqrt{3}(1+12\gamma^2)(\beta'+2\Omega')} \\-144g_{r}(3\lambda-1)\beta'\left[(2\lambda-1)P_{\beta}-(\lambda-1)P_{\Omega}\right]e^{2\sqrt{3}\left[1-4\gamma^2(2\lambda-1)\right]\beta'}\\-72g_{s}(3\lambda-1)^2(2\beta'+\Omega')(2P_{\beta}-P_{\Omega})e^{2\sqrt{3}\left[1-6\gamma^2(3\lambda-1)\right](2\beta'+\Omega')},
    \end{multline}
    \begin{multline}
    \widetilde{V}(\Omega',\beta')=-24(3\lambda-1)e^{-2\sqrt{3}\left[1-2\gamma^2(\lambda-3)\right]\Omega'}+24(3\lambda-1)\Lambda e^{-2\sqrt{3}(1+12\gamma^2)(\beta'+2\Omega')} \\ +12g_{r}(3\lambda-1)e^{2\sqrt{3}\left[1-4\gamma^2(2\lambda-1)\right]\beta'}+12g_{s}(3\lambda-1)e^{2\sqrt{3}\left[1-6\gamma^2(3\lambda-1)\right](2\beta'+\Omega')}.
    \end{multline}

Of course this equation is very complicated to solve in general,
therefore in the next section we will restrict ourselves to the
limiting cases considered in Sec. \ref{S-KSNormal} and look for
analytical solutions.

\section{Infrared and Ultraviolet limits}
\label{S-KSIRUV}

In this section we are going to consider limiting cases of the WDW
equation obtained in the last section. We note that due of the
expansions leading to the expressions
(\ref{ExpansionOmegaM})-(\ref{ExpansionBOP}) we have carried out the
following changes regarding the exponential terms
    \begin{itemize}
        \item $e^{-2\sqrt{3}\Omega}\to e^{-2\sqrt{3}\left[1-2\gamma^2(\lambda-3)\right]\Omega'}$
        \item $\Lambda e^{-2\sqrt{3}(\beta+2\Omega)}\to \Lambda e^{-2\sqrt{3}(1+12\gamma^2)(\beta'+2\Omega')}$
        \item $g_{r}e^{2\sqrt{3}\beta}\to g_{r}e^{2\sqrt{3}\left[1-4\gamma^2(2\lambda-1)\right]\beta'}$
        \item $g_{s}e^{2\sqrt{3}(2\beta+\Omega)}\to g_{s}e^{2\sqrt{3}\left[1-6\gamma^2(3\lambda-1)\right](2\beta+\Omega)}$.
    \end{itemize}
Therefore, we need to be careful in order not to change the signs
of the exponentials so the same terms that we have seen are
contributing in each limiting case in Sec. \ref{S-KSNormal} are not
altered. Therefore, we seek that all the new factors in the
exponentials are positive, that is we demand that
    \begin{equation}\label{ConditionGammaLambda}
    \frac{1}{\gamma^2}>\sup\left(6(3\lambda-1),4(2\lambda-1)\right)= \begin{cases}
    6(3\lambda-1) ,\hspace{0.5cm} \lambda>\frac{1}{5}, \\
    4(2\lambda-1) ,\hspace{0.5cm} \lambda<\frac{1}{5},
    \end{cases}
    \end{equation}
which can be easily fulfilled because $\gamma$ is taken to be a very
small parameter.

As it was precisely the case in Sec. \ref{S-KSNormal}, the IR limit
is obtained when $t>>1$ and $\lambda\to 1$, that is, we look for the
terms containing negative exponentials in the WDW equation.
Therefore, taking a vanishing cosmological constant and a term of
the first form (\ref{AppWF1}) we obtain in this limit
    \begin{equation}\label{WDWGUPIR}
    \left[\left(1+96\sqrt{3}\gamma^2\Omega'e^{-2\sqrt{3}(1+4\gamma^2)\Omega'}\right)\left(P^2_{\beta}-P^2_{\Omega}\right)-48e^{-2\sqrt{3}(1+4\gamma^2)\Omega'}\right]\Psi\left(e^{12\gamma^2}\Omega',\beta'\right) \simeq0 .
    \end{equation}
This result can be compared with the equation discussed in \cite{GUP}.

In the other extreme limit now that we have imposed the conditions
(\ref{ConditionGammaLambda}), we know that the contributing terms
are the $g_{r}$ and $g_{s}$ ones. However, we are going to consider
just the case discussed in Sec. \ref{S-KSNormal} that was shown to
have an analytical solution in the previous case. Therefore we
consider $g_{r}=2$ and $g_{s}=0$. Thus, using an expansion of the
first form (\ref{AppWF11}) for the $g_{r}$ term we obtain the WDW
equation
    \begin{multline}\label{WDWGUPUVO}
    \left\{\left[1+48\sqrt{3}(3\lambda-1)\gamma^2\beta'e^{2\sqrt{3}\left[1-4\gamma^2(2\lambda-1)\right]\beta'}\right]\left[(2\lambda-1)\frac{\partial^2}{\partial\beta'^2}-2(\lambda-1)\frac{\partial}{\partial\beta'}\frac{\partial}{\partial\Omega'}\right.\right. \\ \left.\left. +\frac{\lambda-3}{2}\frac{\partial^2}{\partial\Omega'^2}\right] -36(2\lambda-1)(\lambda-1)\mu^2\gamma^2\beta'e^{2\sqrt{3}\left[1-4\gamma^2(2\lambda-1)\right]\beta'}\frac{\partial}{\partial\Omega'} \right. \\ \left. - 3(2\lambda-1)\mu^2e^{2\sqrt{3}\left[1-4\gamma^2(2\lambda-1)\right]\beta'}\right\}\Psi(\Omega',e^{-12\gamma^2(2\lambda-1)}\beta')\simeq0 .
    \end{multline}
Inspired by the
solution (\ref{SolWDWUV}) we propose a general ansatz of the form
    \begin{equation}\label{UVAnsatz}
    \Psi\left(\Omega',e^{-12\gamma^2(2\lambda-1)}\beta'\right)=e^{\sqrt{3}\nu\Omega'}\left[e^{\sqrt{3}\beta'}\right]^{\frac{\lambda-1}{2\lambda-1}\nu}\chi(\beta') .
    \end{equation}
Substituting back (\ref{UVAnsatz}) into (\ref{WDWGUPUVO}) we obtain
that the $\chi(\beta')$ function obeys the equation
    \begin{equation}\label{GeneralChis}
    \left[\frac{d^2}{d\beta'^2}-V\right]\chi(\beta')=0,
    \end{equation}
where we have defined the function
    \begin{equation}\label{Potentials}
    V=\frac{3\nu^2(3\lambda-1)}{2(2\lambda-1)^2}+\frac{1+12(\lambda-1)\sqrt{3}\nu\gamma^2\beta'}{1+48\sqrt{3}(3\lambda-1)\gamma^2\beta'e^{2\sqrt{3}\left[1-4\gamma^2(2\lambda-1)\right]\beta'}}\frac{24(3\lambda-1)}{(2\lambda-1)}e^{2\sqrt{3}\left[1-4\gamma^2(2\lambda-1)\right]\beta'}.
    \end{equation}
Since we are working in an UV limit, i.e., when
$t=e^{-\sqrt{3}(\beta'+\Omega')}<<1$, we note that this limit can be
achieved by considering $\beta'>>1$. Therefore, we can approximate
the denominator of (\ref{Potentials}) as
    \begin{equation}\label{Approx}
    1+48\sqrt{3}(3\lambda-1)\gamma^2\beta'e^{2\sqrt{3}\left[1-4\gamma^2(2\lambda-1)\right]\beta'}\simeq48\sqrt{3}(3\lambda-1)\gamma^2\beta'e^{2\sqrt{3}\left[1-4\gamma^2(2\lambda-1)\right]\beta'},
    \end{equation}
and then, we can approximate the $V$ function as
    \begin{equation}\label{DefPotentialsUV}
    V^{UV}=V\bigg\rvert_{\beta'>>1}=A_{\nu,\lambda}+\frac{B_{\gamma,\lambda}}{\beta'},
    \end{equation}
where
    \begin{equation}\label{DefsA&B}
        A_{\nu,\lambda}=\frac{3\nu}{2\lambda-1}\left[\frac{\nu(3\lambda-1)}{2(2\lambda-1)}+2(\lambda-1)\right] , \hspace{1cm}
        B_{\gamma,\lambda}=\frac{1}{2\sqrt{3}(2\lambda-1)\gamma^2}.
    \end{equation}

Then, using the approximated potential (\ref{DefPotentialsUV}) into
(\ref{GeneralChis}) we obtain that the general solution is of the
form
    \begin{equation}\label{SolutionChis}
        \chi(\beta')=C M_{\kappa,1/2}\left(2\sqrt{A_{\nu,\lambda}}\beta'\right)+D W_{\kappa,1/2}\left(2\sqrt{A_{\nu,\lambda}}\beta'\right),
    \end{equation}
where $C$ and $D$ are constants, $M_{k,m}(z)$ and $W_{k,m}(z)$ are
the solutions of the Whittaker differential equation and
    \begin{equation}
        \kappa=-\frac{B_{\gamma,\lambda}}{2\sqrt{A_{\nu,\lambda}}}.
    \end{equation}

\begin{figure}[h!]
        \centering
        \includegraphics[width=0.50\textwidth]{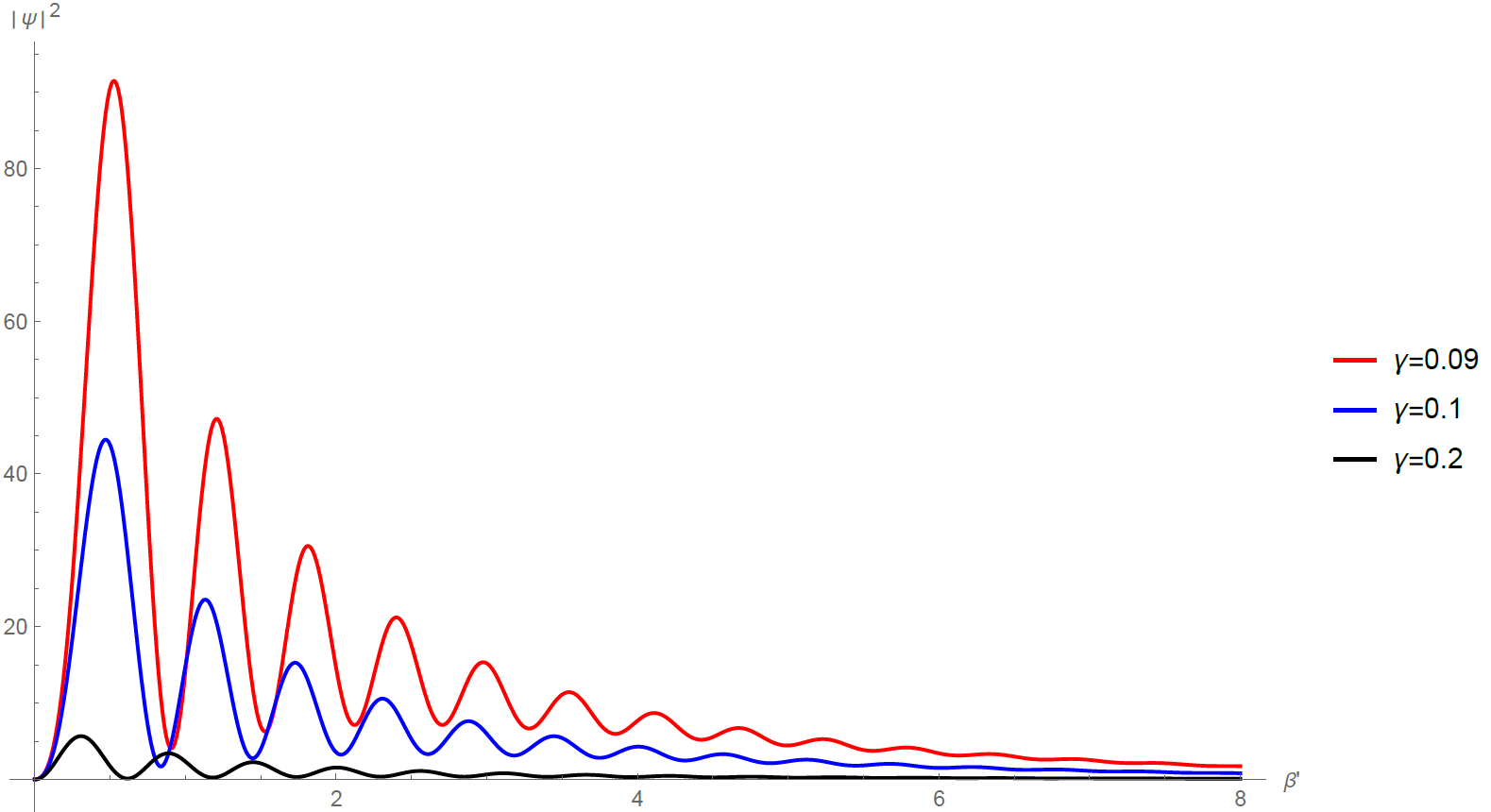}
        \caption{Projection of the absolute value squared of the wave functional $|\Psi|^2$ choosing $|Ce^{\sqrt{3}\nu\Omega'}|^2=1$, $\lambda=1.5$ and $\nu=10i-1$ for $\gamma=0.09$ (red curve), $\gamma=0.1$ (blue curve) and $\gamma=0.2$ (black curve). }
        \label{WFL1.5NC}
    \end{figure}

\begin{figure}[h!]
        \centering
        \includegraphics[width=0.50\textwidth]{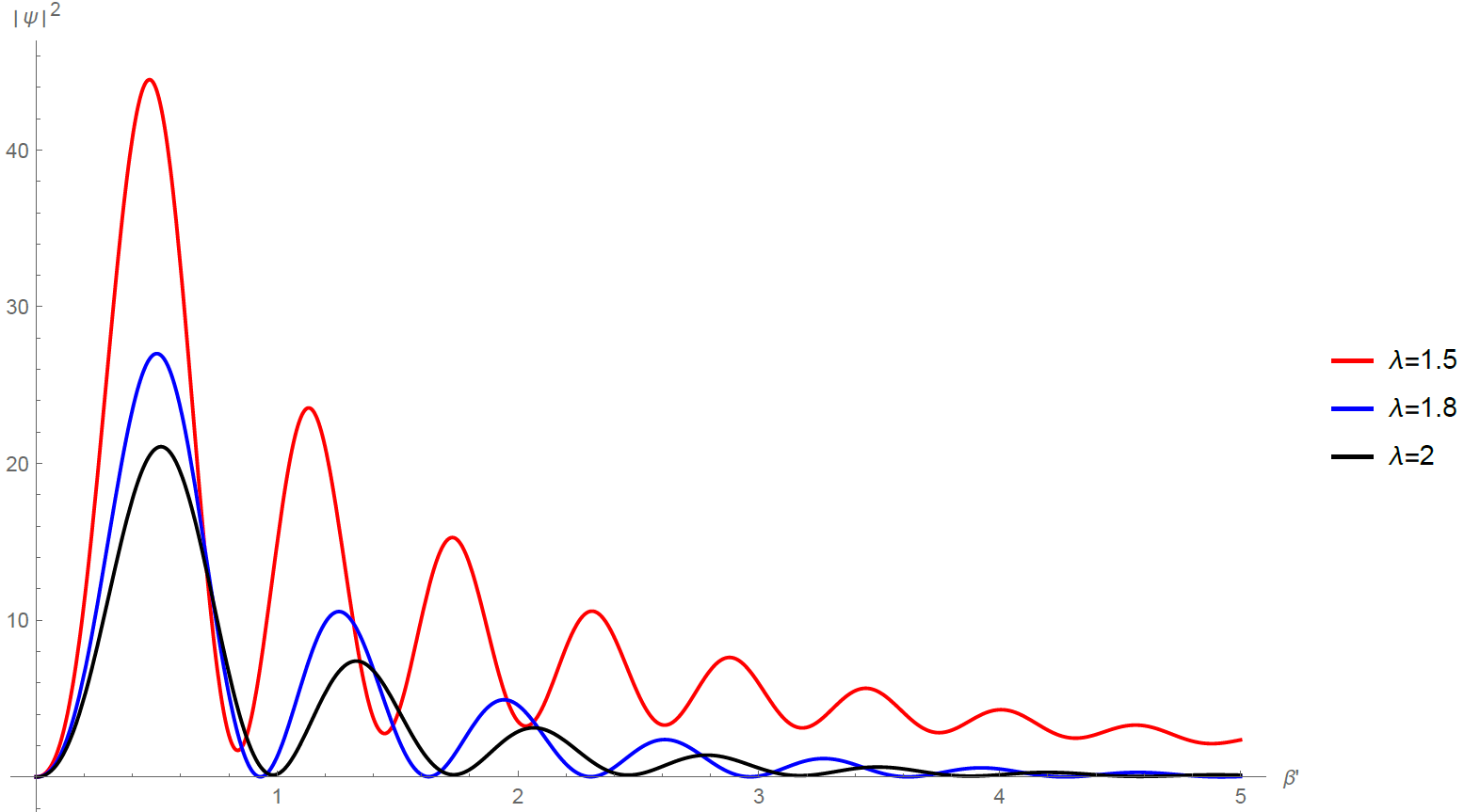}
        \caption{Projection of the absolute value squared of the wave functional $|\Psi|^2$ choosing $|Ce^{\sqrt{3}\nu\Omega'}|^2=1$, $\gamma=0.1$ and $\nu=10i-1$ for $\lambda=1.5$ (red curve), $\lambda=1.8$  (blue curve) and $\lambda=2$  (black curve). }
        \label{WFLVNCF}
    \end{figure}

\begin{figure}[h!]
        \centering
        \includegraphics[width=0.50\textwidth]{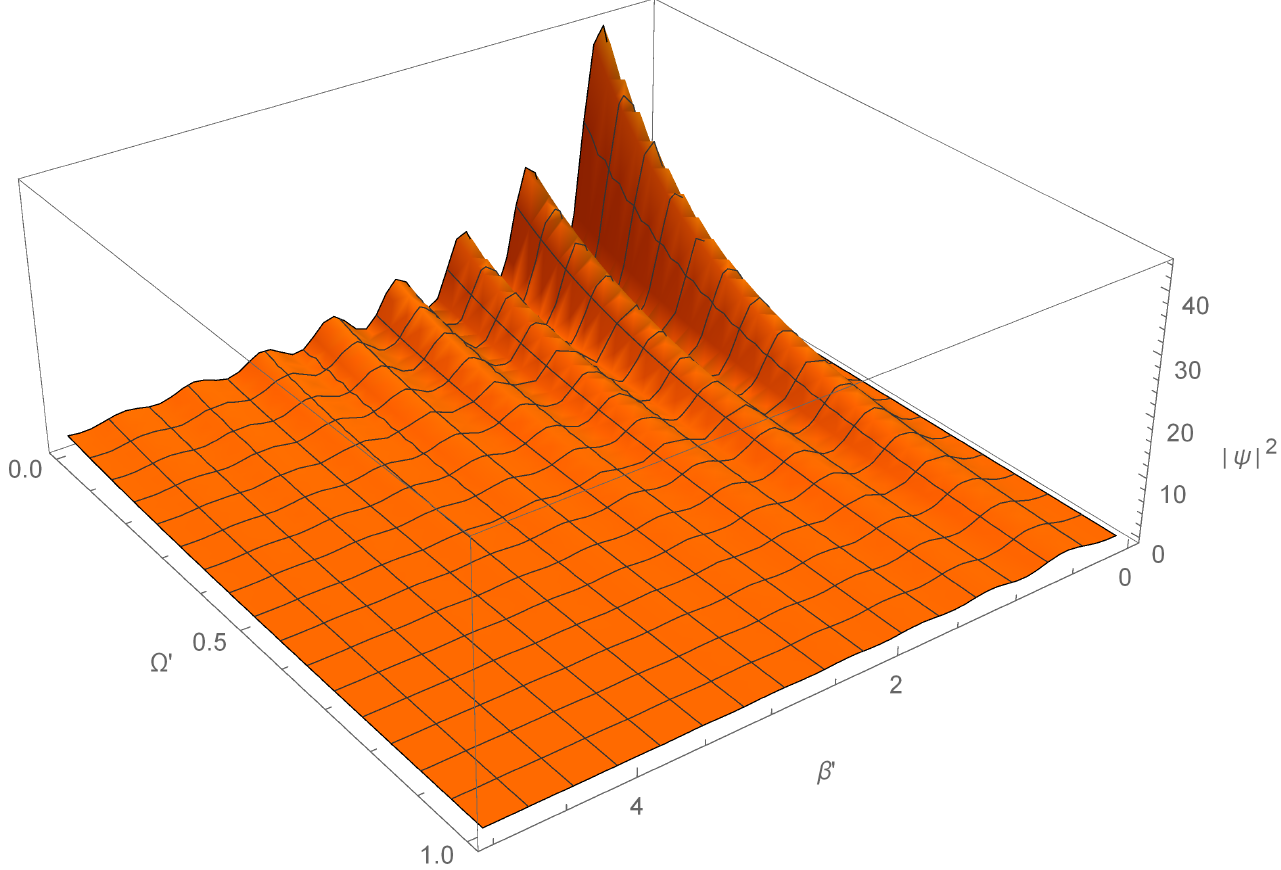}
        \caption{Absolute value squared of the wave functional $|\Psi|^2$ choosing $|C|^2=1$, $\lambda=1.5$, $\gamma=0.1$ and $\nu=10i-1$. }
        \label{WF3D}
    \end{figure}

We note that contrary to the situation discussed in Sec.
\ref{S-KSNormal}, in the present case there is no restriction that
lead us to take values of $\nu$ purely imaginary. Therefore in this
case we have the freedom to propose values of $\nu$ real, purely
imaginary or even complex focusing only on obtaining a well defined
behaviour for the wave functionals. However, as it is usual in quantum
cosmological models, we are looking for oscillatory solutions which
can be interpreted as Lorentzian geometries
\cite{Hawking,Wada:1985cp,Halliwell:1989myn} and it can be proved \cite{OscillationsN,OscillationsT} that the Whittaker equation found for the function $\chi$ does not have oscillatory solutions if $A_{\nu,\lambda}$ is a positive real number and $B_{\gamma,\lambda}$ is also positive. This can be achieved if $\nu$ takes positive real values and $\lambda>1/2$ (which is the values of interest to us). Therefore, positive real numbers for $\nu$ will not be considered in the following. We note that in all cases the
function $|W_{\kappa,1/2}|^2$ exhibits a monotonically decreasing
behaviour and therefore we can consider in (\ref{SolutionChis}) only
the $M$ function, therefore we obtain finally that the wave
functional is given by
    \begin{equation}\label{SolWaveFunc}
        \Psi\left(\Omega',e^{-12\gamma^2(2\lambda-1)}\beta'\right)=Ce^{\sqrt{3}\nu\Omega'}\left[e^{\sqrt{3}\beta'}\right]^{\frac{\lambda-1}{2\lambda-1}\nu}M_{\kappa,1/2}\left(2\sqrt{A_{\nu,\lambda}}\beta'\right).
    \end{equation}
However, the behaviour of the squared absolute value of this wave
functional $|\Psi|^2$ with respect to $\beta'$ is troublesome. In
most cases it only acts as a monotonically increasing function which
is not a behaviour we are interested in. Nonetheless, for some values
of the parameters it is possible to obtain an oscillatory behaviour.
For example for values of $\lambda>1$ it is possible to obtain a
decreasing oscillatory behaviour if we choose $\nu$ as a negative
real number (with $A_{\nu,\lambda}$ nonpositive) or a complex number with negative real part. In Figures
\ref{WFL1.5NC} and \ref{WFLVNCF} we show plots of the projection of
$|\Psi|^2$ with respect to $\beta'$ choosing
$|Ce^{\sqrt{3}\nu\Omega'}|^2=1$. In Figure \ref{WFL1.5NC} we choose
$\lambda=1.5$ and $\nu=10i-1$, we show curves for three different
values of $\gamma$. We note that as $\gamma$ increases, the absolute
value squared of the wave functional decreases. In the other hand,
in Figure \ref{WFLVNCF} we also choose  $\nu=10i-1$ and
$\gamma=0.1$, however in this case  we show curves for three
different values of $\lambda$. We obtain also that as $\lambda$
increases, the absolute value squared of the wave functional
decreases. We note however that in both cases we are limited to a
small interval to vary both parameters $\gamma$ or $\lambda$, since
if we change one of them too much without changing the other or
$\nu$, the behaviour can be easily spoiled. We also note that the
oscillatory behaviour is present at first but it is reduced as
$\beta'$ increases and eventually it behaves just as a monotonically
decreasing function, thus the parameters must be chosen such that
the region of oscillatory behaviour is in agreement with the region
where the approximation (\ref{Approx}) is valid. This also happens
if $\nu$ is taken as a real negative number. Finally, in Figure
\ref{WF3D} we show the absolute value squared of the complete wave
functional choosing $|C|^2=1$, $\nu=10i-1$, $\lambda=1.5$ and
$\gamma=0.1$. We see that the dependence on $\Omega'$ is just a
negative exponential as shown in (\ref{SolWaveFunc}). 


\begin{figure}[h!]
        \centering
        \includegraphics[width=0.50\textwidth]{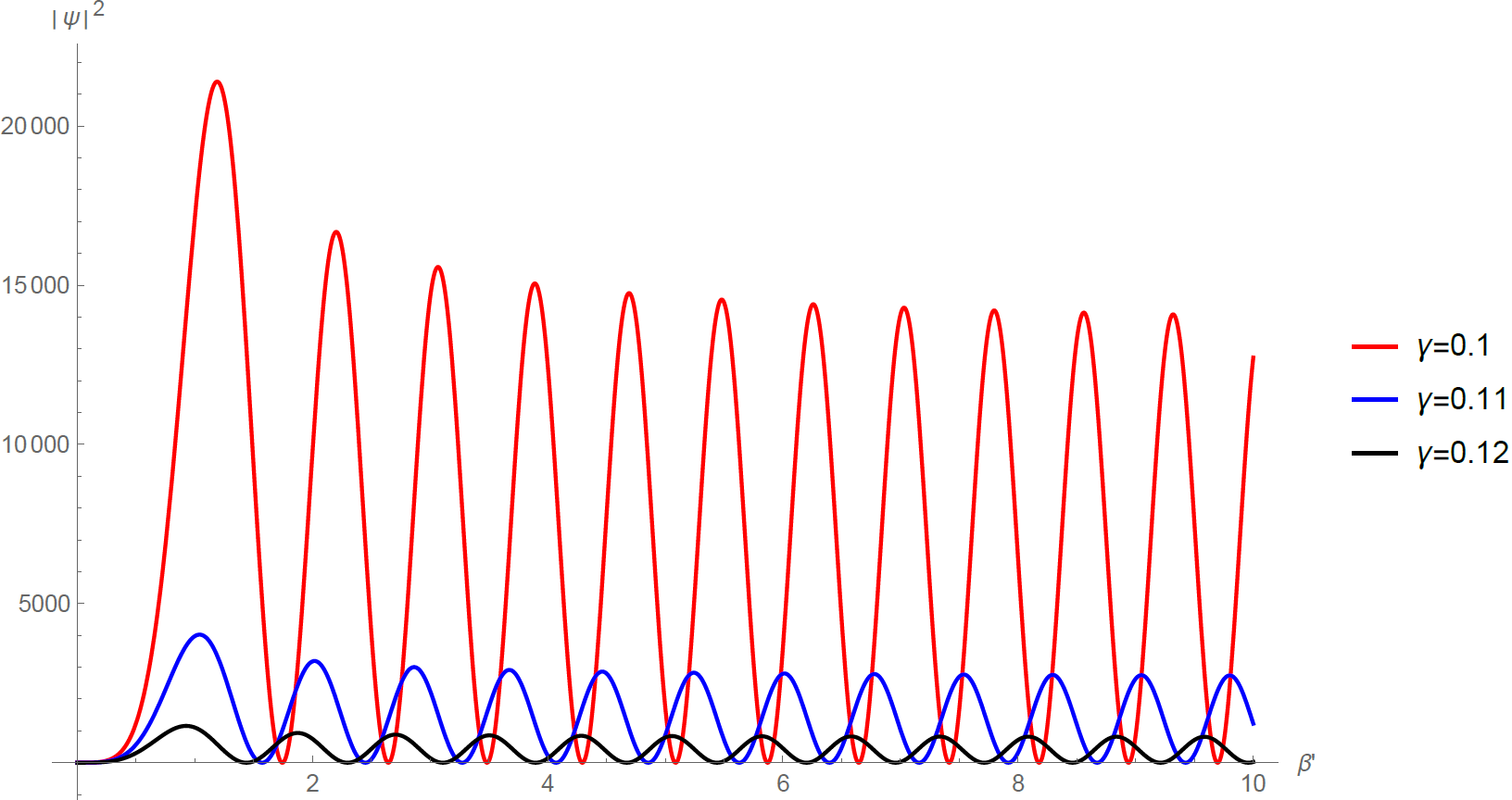}
        \caption{Projection of the absolute value squared of the wave functional $|\Psi|^2$ choosing $|Ce^{\sqrt{3}\nu\Omega'}|^2=1$, $\lambda=1$ and $\nu=5i$ for $\gamma=0.1$ (red curve), $\gamma=0.11$ (blue curve) and $\gamma=0.12$ (black curve). }
        \label{WFL1NI}
    \end{figure}

It can be appreciated from (\ref{DefsA&B}) that the case in which $\lambda=1$
is a particular one since in that case $A_{\nu,\lambda}$ only
depends on $\nu^2$. For this particular case an oscillatory solution
can also be found if we choose $\nu$ as a completely imaginary
number, but in this case the oscillations do not decrease in height
after a few peaks. In Figure \ref{WFL1NI} we show the plot of the
projection of $|\Psi|^2$ with respect to $\beta'$ choosing
$|Ce^{\sqrt{3}\nu\Omega'}|^2=1$ in this particular case. We take
$\nu=5i$ and show curves for three different values of $\gamma$. In
this case we also have the result found previously, that is, the
absolute squared value of the wave functional increases inversely
with $\gamma$.

From this analysis we can conclude that the form of the solution considering a GUP (\ref{UVAnsatz}) has a similar form as the one obtained with the usual uncertainty relation (\ref{SolWDWUV}) but the special function is modified from a Bessel function to a Whitakker function. This of course leads us to a different behaviour for the absolute value of the squared wave functional. For $\lambda>1$ we have in both cases a finite region of oscillatory behaviour starting from the origin at $\beta=0$ or $\beta'=0$. However in the solution (\ref{SolWDWUV}) this is achieved when $\nu$ is a purely imaginary number and the height of the oscillations increases until it reaches a maximum value whereas after considering a GUP the height of the oscillations have the opposite behaviour, that is they decrease until the oscillatory behaviour stops and it can be found when $\nu$ is a real negative number or a complex number with negative real part. We also note that in the GUP case we used a limit of $\beta'>>1$ and therefore some of the region of oscillatory behaviour is not accessible. In both cases we obtain that as $\lambda$ increases the squared wave functional decreases. Finally, the $\lambda=1$ case is a special one in the GUP scenario where we obtain a oscillatory behaviour that does not decrease in height, whereas in the solution (\ref{SolWDWUV}) this value of $\lambda$ is not special, it produces the same behaviour described earlier.

\textbf{\textit{The $\boldsymbol{\gamma\to0}$ limit}} \\
Before we end up let us discuss a subtle issue regarding the limit $\gamma\to0$. If we take such limit in Eq. (\ref{WDWGUPUVO}) it simplifies yielding (\ref{WDWOUV}) as expected. We also note that in this limit Eq. (\ref{GeneralChis}) gives the equation obeyed by the modified Bessel function of second order with the appropiate change of variables. Therefore in this limit the ansatz (\ref{UVAnsatz}) gives smoothly the same solution (\ref{SolWDWUV}) as expected. However, we note from (\ref{DefsA&B}) that in this limit $B_{\gamma,\lambda}$ is divergent and therefore we cannot obtain the solution in the standard case from the solutions found by considerind a GUP (\ref{SolWaveFunc}). This is because the ultraviolet approximation (\ref{Approx}) considered in order to obtain analytic solutions is no longer valid if we take the $\gamma\to0$ limit. Therefore, the region of validity of the  solutions found in terms of the Wittaker functions depend not only on $\beta'$ but also in $\gamma$ as well.

We may be tempted to look for analytic solutions that smoothly recover the solution (\ref{SolWDWUV}) by expanding the denominator in the second term of (\ref{Potentials}) and keep only up to second order in the parameter $\gamma$. However, we note that in this case $\gamma$ is acompannied by a positive exponential term depending on $\beta'$ and therefore it is not justified to keep only up to second order on $\gamma$ since in this case the whole term is not small in general. In fact, it can only be small for small values of $\beta'$ which is the opposite of the UV limit considered. Therefore we believe it is more appropriate to use the UV approximation in the denominator and restrict ourselves to its corresponding region of validity.

\section{Final Remarks}
\label{S-FinalR} In the present article we have studied the
Wheeler-DeWitt equation coming from a Kantowski-Sachs metric in
Ho\v{r}ava-Lifshitz gravity with coordinates of minisuperspace that
obey a GUP of the form proposed in \cite{Kempf:1994su} using the
procedure presented originally in \cite{GUP}.

We first presented the WDW equation coming from the coordinates that
obey the usual uncertainty relation and considered two limiting
cases. In the IR limit the equation was reduced to the one found if
one uses GR \cite{GUP}. On the other hand, in the UV limit we were
able to find the equation coming from HL gravity with the
projectability condition and with detailed balance, but in this case
it was found to be only an approximation of the theory without
detailed balance after choosing specific values for the $g_{r}$ and
$g_{s}$ constants. In both cases analytic solutions were found and the behaviour of the UV solution was presented in Figure \ref{WGUP}. We found that the solutions describe a finite region of oscillatory behaviour with an increasing height until it reaches a maximum peak, then the oscillatory behaviour stops and it starts decreasing. We also found that as $\lambda$ increases the absolute value squared of the wave functional decreases.

We then moved on to study the WDW equation when the coordinates on
minisuperspace obey a GUP. We used the general procedure to modify
the WDW equation obtained previously by using the relations between
both sets of coordinates and consider only terms up to second order
in momentum as well as second order on the GUP parameter $\gamma$.
Since all terms contain exponentials of the form a coordinate
times its corresponding momentum, we found that there was two
possible procedures to obtain the WDW equation depending on how we
take this term when acting on the wave functional. They can be
interpreted as terms that scale the coordinates in the wave
functional as in \cite{GUP}, this interpretation was used when we
studied the limiting cases and look for analytic solutions. However,
they can also be interpreted as to contribute with terms of linear
momenta without rescaling the coordinates on the wave functional,
this form was useful to obtain the WDW equation in general.

Since the general WDW equation, which is valid in all cases, was found to be very
difficult, we proceeded to study the limiting cases considered
before. In the IR limit we found an equation that can be compared to
the result obtained by using GR in \cite{GUP}. In the UV limit we
considered special values of $g_{r}$ and $g_{s}$ as we
mentioned before. We showed that this limit could be achieved if we take
$\beta'>>1$ and therefore, we approximated the resulting equation.
We then obtained an analytic solution of such equation in terms of
special functions. We found that the general form of the solution
was the same as the one encountered when the coordinates obey the
standard uncertainty principle but now the special function was not
a Bessel function of the second kind, it was instead a Whittaker
function. We also noted that contrary to the standard case (without considering a GUP) in this case the parameter $\nu$ in the
ansatz was not subject to a restriction and therefore we were free
to choose $\nu$ as a real, purely imaginary or even a complex
number. We only focused on the values that leaded us to a correct
behaviour for the resulting wave functionals. We found  that only one
of the two possible Whittaker functions was able to produce
oscillatory solutions but the dependence on the parameters  $\nu$,
$\gamma$ and $\lambda$ was found to be very restringing and the
oscillatory behaviour is not present for any arbitrary values. It was also found that this behaviour was encountered only when $\nu$ was not allowed to be a positive real number. For
$\lambda>1$ plots of the squared value of the wave functional
$|\Psi|^2$ were presented in Figures \ref{WFL1.5NC}, \ref{WFLVNCF}
and \ref{WF3D}. We noted that the oscillatory behaviour is strong at
first but then it decreases until it disappears, therefore it is
necessary to look for an interval where the oscillatory behaviour and
the approximation used are both valid. The peaks of the oscillations were found to decrease in height contrary to the behaviour found for the case without considering a GUP. We also showed that the value
of $|\Psi|^2$ decreases if $\gamma$ or $\lambda$ was increased. The
case $\lambda=1$ was found to be an special one, for this case an
oscillatory behaviour could also be found and we presented a plot of
it in Figure \ref{WFL1NI}, however in this case the height of the
oscillations does not decrease after a few peaks. In this case
it was also shown that $|\Psi|^2$ decreases as $\gamma$ increases. Contrary to the standard scenario without a GUP where this case is not special and the behaviour is the same as for any other value of $\lambda$.

Finally, it would be interesting to consider a noncommutative
deformation of the general WDW equation (\ref{WDWO}) found in this
article. It is well known that the WDW equation (\ref{WDWOGR}) in
GR, can be noncommutatively deformed in its Heisenberg algebra through
a parameter $\theta$ \cite{Garcia-Compean:2001jxk}. This
noncommutative deformation is also relevant at high energies and
this represents another UV effect that should be taken	 into account
since it is also related to the existence of a minimal length. Some
solutions of the noncommutative deformation of (\ref{WDWOGR}) are
known \cite{Garcia-Compean:2001jxk}. Thus they would be useful to
find new solutions to the noncommutative deformation of the general
WDW equation (\ref{WDWO}) similar to the procedure carried out in the present paper.
Some of this work is in progress and will be reported elsewhere.

 \vspace{1cm}
\centerline{\bf Acknowledgments} \vspace{.5cm} It is a pleasure to thank Prof. O. Obreg\'on for comments on the manuscript. D. Mata-Pacheco would also like to thank CONACyT for a grant.



\end{document}